\documentclass[journal=jacsat,manuscript=article]{achemso}

\usepackage[version=3]{mhchem} % Formula subscripts using \ce{}
\usepackage{multirow}
\usepackage{makecell}
\usepackage{cellspace}
\usepackage{booktabs}
\author{Janghoon Ock}
\affiliation[ChemE]
{Department of Chemical Engineering, Carnegie Mellon University, 5000 Forbes Street, Pittsburgh, PA 15213, USA}

\author{Parisa Mollaei}
\author{Amir Barati Farimani}
\email{barati@cmu.edu}
\affiliation[MechE]
{Department of Mechanical Engineering, Carnegie Mellon University, 5000 Forbes Street, Pittsburgh, PA 15213, USA}

\title[An \textsf{achemso} demo]
  {GradNav: Accelerated Exploration of Potential Energy Surfaces with Gradient-Based Navigation}

\abbreviations{IR,NMR,UV}

\begin{document}

\begin{tocentry}

\centering
\includegraphics{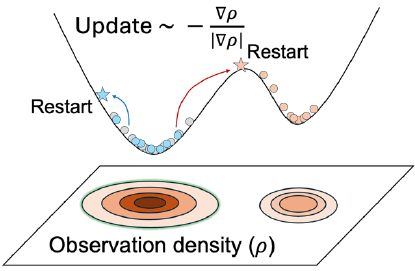} 
\label{fig:toc}

\end{tocentry}

\begin{abstract}
The exploration of molecular systems' potential energy surface is important for comprehending their complex behaviors, particularly through identifying various metastable states. However, the transition between these states is often hindered by substantial energy barriers, demanding prolonged molecular simulations that consume considerable computational efforts. Our study introduces the GradNav algorithm, which enhances the exploration of the energy surface, accelerating the reconstruction of the potential energy surface (PES). This algorithm employs a strategy of initiating short simulation runs from updated starting points, derived from prior observations, to effectively navigate across potential barriers and explore new regions. To evaluate GradNav's performance, we introduce two metrics: the deepest well escape frame (DWEF) and the search success initialization ratio (SSIR). Through applications on Langevin dynamics within Müller-type potential energy surfaces and molecular dynamics simulations of the Fs-Peptide protein, these metrics demonstrate GradNav's enhanced ability to escape deep energy wells, as shown by reduced DWEF values, and its reduced reliance on initial conditions, highlighted by increased SSIR values. Consequently, this improved exploration capability enables more precise energy estimations from simulation trajectories.

\end{abstract}

% For example, where Langevin dynamics struggle to escape the deepest well within 150,000 frames (DWEF > 150k), GradNav achieves this in just 700 frames (DWEF = 700), under identical physical conditions. Furthermore, while Langevin dynamics capture only half of the potential wells with multiple initial points across the landscape (SSIR = 50\%), GradNav successfully identifies all of them (SSIR = 100\%), showcasing its superior exploratory capability and efficiency.

\section{Introduction}

The reconstruction of potential energy surface (PES) based on molecular simulation is important for modeling and understanding molecular systems, especially proteins. Exploring this energy surface through molecular simulations enables an understanding of the intricate behaviors of complex molecular systems. A crucial aspect of these explorations is the identification of multiple metastable states, which manifest as local minima or potential wells in the energy surface\cite{Wales2006}. These states are essential for gaining insights into the characteristics of atomic and molecular systems. For instance, the transition between folded and unfolded states, representing metastable states in proteins, significantly alters the functional attributes of protein molecules\cite{Dobson2003, protein_folding, Lemcke2023, mollaei2023transition}. Additionally, identifying the boundaries of metastable states in superheated crystals and supercooled liquids plays a crucial role in understanding nucleation behavior \cite{metastable_crystal, Sciortino_2005}. All-atom molecular simulations, however, often face a significant challenge: they tend to get trapped in deep potential wells of lower energy states\cite{Yang2019, okamoto2009generalized}. This entrapment necessitates excessively long simulation trajectories to thoroughly escape these wells and continue exploration across the energy landscape. Such prolonged simulations demand substantial computational resources, rendering them less feasible for many applications.

To address these challenges, numerous enhanced sampling methods have been introduced\cite{Yang2019}. These methods typically modify the physical parameters of the simulation to facilitate the exploration of the energy surface, such as by applying bias potentials or elevating the temperature. For example, bias potentials are utilized in metadynamics\cite{metadynamics1, metadynamics2}, umbrella sampling\cite{umberella}, and variationally enhanced sampling (VES)\cite{ves} to help the system escape from potential energy wells and sample a broader range of configurations. Techniques such as replica exchange molecular dynamics (REMD)\cite{remd} and temperature-accelerated dynamics (TAD)\cite{temp-accel-dynamics, temp-accel-md} increase the temperature of the simulation to accelerate the occurrence of rare events, thereby enabling the study of processes that occur over longer timescales. However, the introduction of bias or temperature perturbations requires careful correction to recover unbiased physical properties, as these modifications can alter the fundamental nature of the events being studied or lead to different mechanisms being observed\cite{bias_effect, bias_effect2}.

In response to these limitations, we propose an observation-driven algorithm designed to accelerate the exploration of molecular systems while adhering strictly to the original physics of the system. This algorithm leverages the concept of restarting short chunks of simulations, guided by previous observations, to systematically optimize the starting point away from previously explored regions. By updating the initial points based on the gradient of observation density, this method systematically directs the exploration away from regions with high observation concentration. This facilitates quicker escape from deep potential wells, thereby enabling the investigation of nearby regions for the discovery of additional metastable states. Importantly, this methodology preserves the original physical settings of the system, guaranteeing the authenticity of each simulation chunk and thus ensuring a reliable investigation of the energy surface. By relying exclusively on observations from previous simulations to guide the exploration towards unexplored states, without imposing any artificial biases to the physical settings, our approach offers a cost-effective and physically consistent strategy for navigating the energy surface.

Moreover, a diverse range of machine learning techniques, such as graph neural networks, transformers, and multimodal models, is being increasingly utilized for modeling atomic systems\cite{mlp, ml_mat, Wang2023, catberta, ock2024multimodal}. For example, autoencoders have demonstrated success in translating the Brownian dynamics trajectories of a 2D energy surface into a latent space representation\cite{Ramil2022}. Additionally, supervised machine learning has been employed to pinpoint suitable collective variables for study\cite{autocv}. Furthermore, flow-based Boltzmann generators have effectively mapped the structure of the bovine pancreatic trypsin inhibitor (BPTI) protein into a latent space, while retaining distributional characteristics from the original space\cite{bg_noe}. Successful mapping of atomic systems into latent space preserves their real-space distribution. Hence, areas densely populated in real space should correspond to similarly dense regions in latent space. This indicates the feasibility of applying observation-driven exploration within latent space.

% contributions
This study presents the observation density gradient-based navigation (GradNav) algorithm, designed to enhance the exploration of potential energy surfaces by facilitating the escape from deep potential wells. The algorithm is designed to accelerate the exploration process across molecular energy surfaces while maintaining the original physical parameters of molecular simulations—such as potential energy and temperature—unchanged. This approach guarantees that the resulting trajectory is a true reflection of the system's inherent physical behavior. Furthermore, we introduce two metrics to evaluate the algorithm's ability to escape from deep potential wells and its robustness against variations in initialization points. Utilizing these metrics, we assess the algorithm's effectiveness with model systems under Langevin dynamics within both M\"uller potential energy surface and its modified version\cite{silvia2017}. The algorithm's validity is further confirmed through its application to a real-world example: the molecular dynamics simulation of the Fs-Peptide protein. The trajectories in the dataset provide valuable insights into the folding dynamics of the protein, which can aid in understanding protein folding mechanisms\cite{fs-peptide, mollaei2023unveiling, peptide_bert}.

\section{Results and discussion}
\subsection{Framework}

\begin{figure*}[htbp] %[h!]
\centering
\includegraphics[width=0.8\textwidth]{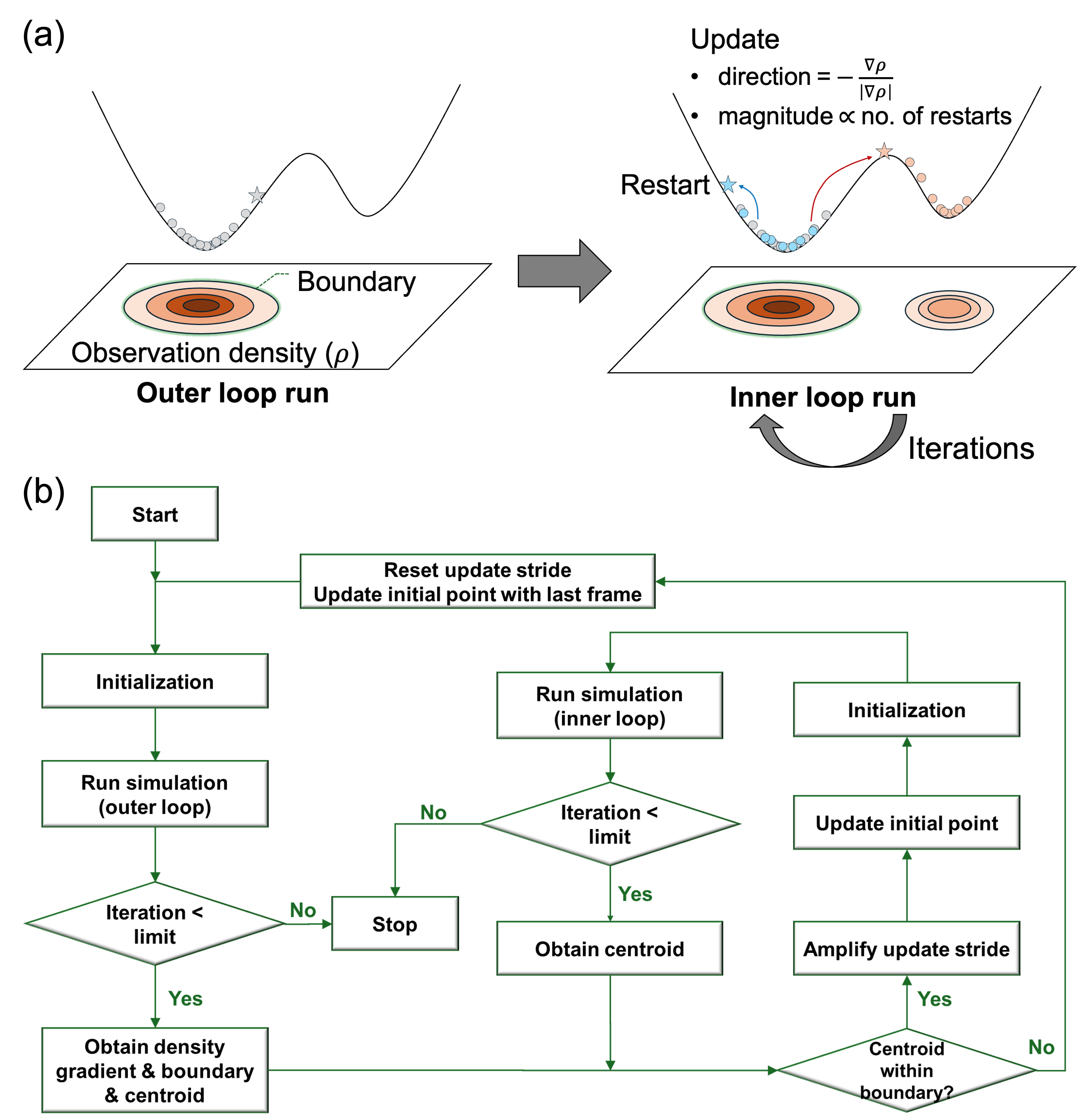} 
\caption{Overview of the GradNav Algorithm. \textbf{a} Illustration of the update rule: the update rate is increased if the centroid of the subsequent trajectory falls within the previously determined boundary. This process repeats until a new potential well is identified. \textbf{b} Flowchart detailing the steps of the GradNav algorithm.}
\label{fig:framework}
\end{figure*}

The observation density gradient-based navigation (GradNav) algorithm introduces a data-driven method for effectively navigating the potential energy surface, specifically designed to address the common challenge in molecular simulations: escaping deep potential wells once trapped. Unlike ordinary molecular simulations that rely on prolonged simulation runs to escape potential wells, this approach employs repeated initiations of molecular simulations to facilitate a more dynamic exploration of the energy surface. Utilizing the observation density gradient, this method intelligently directs simulations toward less explored, potentially more insightful areas. As illustrated in Figure \ref{fig:framework}, the algorithm incorporates two iterative loops of molecular simulation: the outer loop runs molecular simulations with relatively long frames to calculate the observation density gradient and determine the boundaries of potential wells; the inner loop conducts shorter, exploratory simulations to discover new areas without using excessive computational power. Although the duration of the outer loop simulations exceeds that of the inner loop ones, these frames are considerably shorter compared to conventional extended molecular simulations.

The algorithm starts with a molecular simulation in an outer loop, which runs for a relatively long frame sequence, often exceeding 100 frames. This phase aims to accumulate sufficient trajectory data for calculating observation density, spatial boundaries of observations, and a centroid representing their average location. In this study, we designated 300 to 500 frames for this purpose. While this number varies by system, exceeding 100 frames generally suffices for these calculations. The boundary defined in the outer loop run is used to determine if subsequent simulations manage to identify a new potential well or a metastable state. 

Upon the centroid's placement within this boundary, the algorithm's inner loop is activated. This indicates a failure to detect multiple potential wells, thereby necessitating the initiation of a new simulation for further exploration. The inner loop begins with molecular simulations of excessively short duration. These simulations are conducted iteratively, with an update rate that gradually escalates, until the centroid of a subsequent run extends past the boundary defined by the preceding outer loop run. The update of the initialization point of the simulation is defined as follows: 

\begin{subequations}
\begin{align}
x_{n+1}^i &= x_n^l - \frac{\beta (1+v_n/k)}{\lvert \nabla \rho \rvert} \nabla \rho, \\
v_{n+1} &= \gamma (v_n + \gamma)
\end{align}
\end{subequations}

A new initialization point \(x^i_{n+1}\) is determined using the final point from the preceding run \(x^l_{n}\), by applying the update rate -\(\frac{\beta (1+v_n/k)}{\lvert \nabla \rho \rvert} \nabla \rho\). Here, \(\rho\) denotes the observation density, \(\beta\) and \(k\) serve as hyperparameters, and \(v_n\) represents the update stride at iteration \(n\), which regulates the progressive intensification of the iteration. \(\gamma\) acts as a conditional parameter, becoming one if the centroid lies within the boundary, and zero otherwise, effectively resetting the update rate to zero. The initial point updates are intentionally directed from highly populated regions towards areas with lower population densities, aligning with the direction of the negative observation density gradient ($\nabla \rho$). The magnitude of these updates increases linearly with the number of inner loop iterations, representing the attempts to identify other metastable states. If the centroid remains within the initial boundary, the update stride is incrementally increased by one, linearly raising the update magnitude. Upon the centroid successfully moving beyond the boundary, the update step is reset to zero, allowing a new outer loop simulation and exploration cycle to commence. Simulations in both the inner and outer loops proceed over a predetermined number of frames. The hyperparameters \(\beta\) and \(k\) govern the increase of the update stride, influencing the linear increase's intercept and slope, respectively. The specific hyperparameters used in each case are detailed in Table \ref{tab:parameter}. It should be noted that the hyperparameters have not been optimized for maximal exploration efficiency, as the primary objective of this paper is to illustrate the concept of the newly proposed algorithm.

% maybe add more explanation about hyperparameters

This iterative re-initiation enables the GradNav algorithm to explore the potential energy surface efficiently without imposing any artificial bias on the simulations. Hence, each simulation run reflects the outcomes of the precise physical settings. Additionally, by collecting the starting points of each run, it is possible to determine the moment of transition to another metastable state.

\begin{table}[ht]
\centering
\caption{Hyperparameters and simulation settings for different systems}
\label{tab:parameter}
\begin{tabular}{lccc}
\toprule
Parameter & Müller & Modified Müller & Fs-Peptide \\
\midrule
$\beta$ & 0.75 & 1 & 0.1 \\
$k$ & 100 & 20 & 100 \\
Outer loop frames & 500 & 500 & 300 \\
Inner loop frames & 50 & 50 & 40 \\
Total iteration number & 10,000 & 10,000 & 10,000 \\
\bottomrule
\end{tabular}
\end{table}

\subsection{Escaping the Deep Potential Well}

The GradNav algorithm is primarily designed to expedite the escape of molecular simulations from deep potential wells effectively. To quantify this escape capability, we introduce a metric denoted as the deepest well escape frame (DWEF), which measures the number of frames required for the simulation to exit the deepest potential well. We conduct comparisons using Langevin dynamics simulations of a single particle within the Müller potential and its modified versions. The Müller potential is a widely recognized potential energy surface, extensively studied for its characteristics, especially in evaluating the efficacy of algorithms designed to identify reaction paths or metastable states \cite{Ramil2022, silvia2017, bg_noe}. It features three minima: two are relatively deep, and one, located between the two, is comparatively shallow. Modifications to the Müller potential include creating a deeper valley amidst two comparatively shallow metastable states\cite{silvia2017}, aiming to test the algorithm's ability to navigate across the deepest valley to reach these shallower wells. Further details on the potential energy surface are provided in the Methods section.

\begin{figure*}[htbp] %[h!]
\centering
\includegraphics[width=0.8\textwidth]{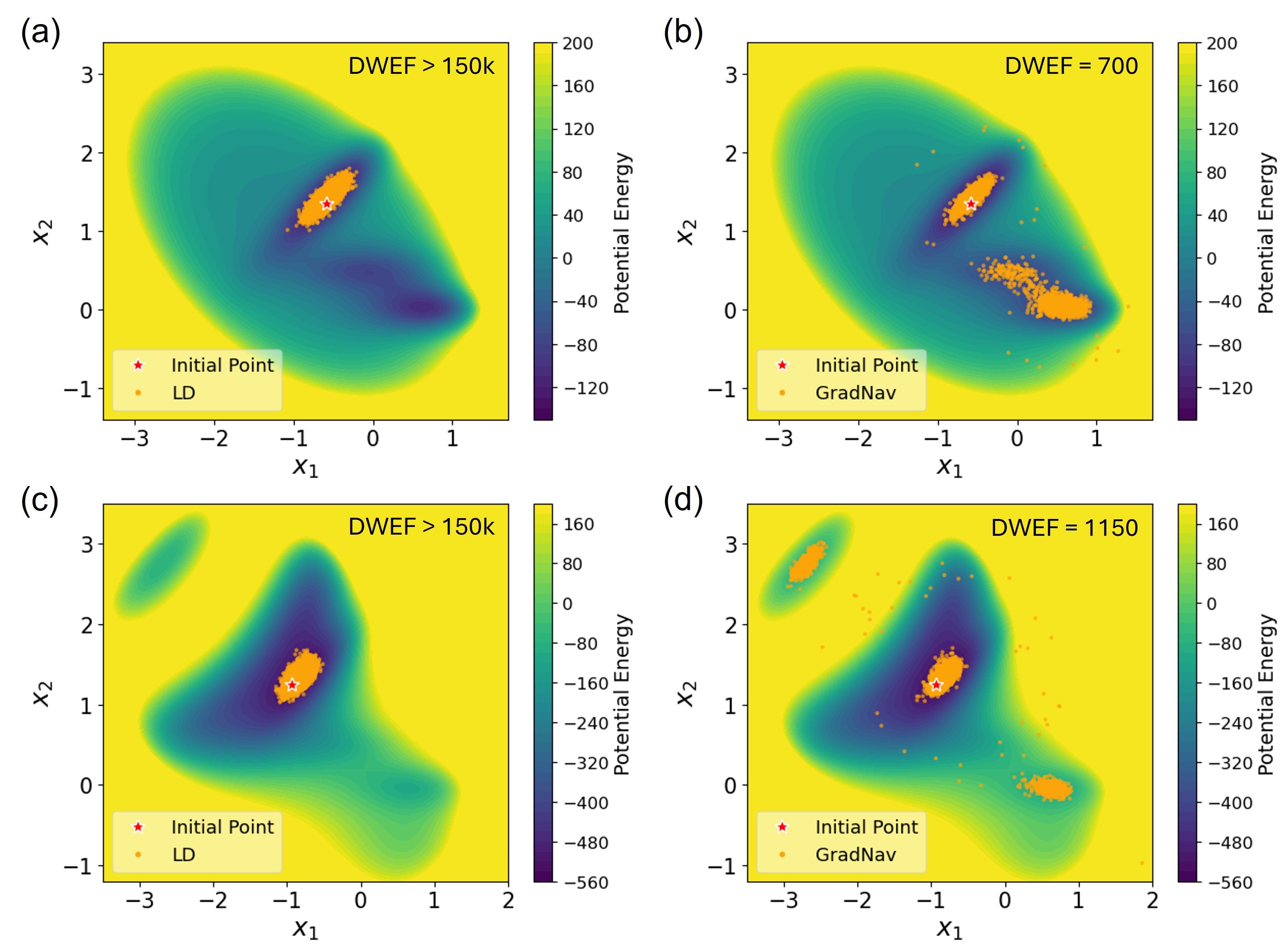} 
\caption{Trajectories generated using Langevin Dynamics (LD) and the GradNav algorithm, starting from the deepest valley.}
\label{fig:mullers_traj}
\end{figure*}

% illustrate simulation settings
Simulations initiate from the initial positions situated at the deepest potential wells within each energy surface. As these simulations progress, we track the number of frames needed to escape these wells. The trajectories generated from both Langevin dynamics simulations and the application of the GradNav algorithm are depicted in Figure \ref{fig:mullers_traj}. Ordinary Langevin dynamics simulations, extending over 150,000 frames, fail to exit the deepest wells in both the Müller and the modified Müller potential energy surfaces, resulting in a DWEF count exceeding 150,000, as shown in Figure \ref{fig:mullers_traj} (a) and (c). Conversely, with the GradNav algorithm, escapes from the deepest wells are achieved within merely 700 and 1,150 frames for the Müller and modified Müller potential energy surfaces, respectively, as illustrated in Figure \ref{fig:mullers_traj} (b) and (d). A lower DWEF suggests that the simulation is capable of escaping the potential well, thereby enabling continued exploration across the energy surface. Additionally, Figure S1 in the Supporting Information depicts trajectories exclusively gathered from the outer loop and the initial points for each inner loop run.

\begin{figure*}[htbp] %[h!]
\centering
\includegraphics[width=0.8\textwidth]{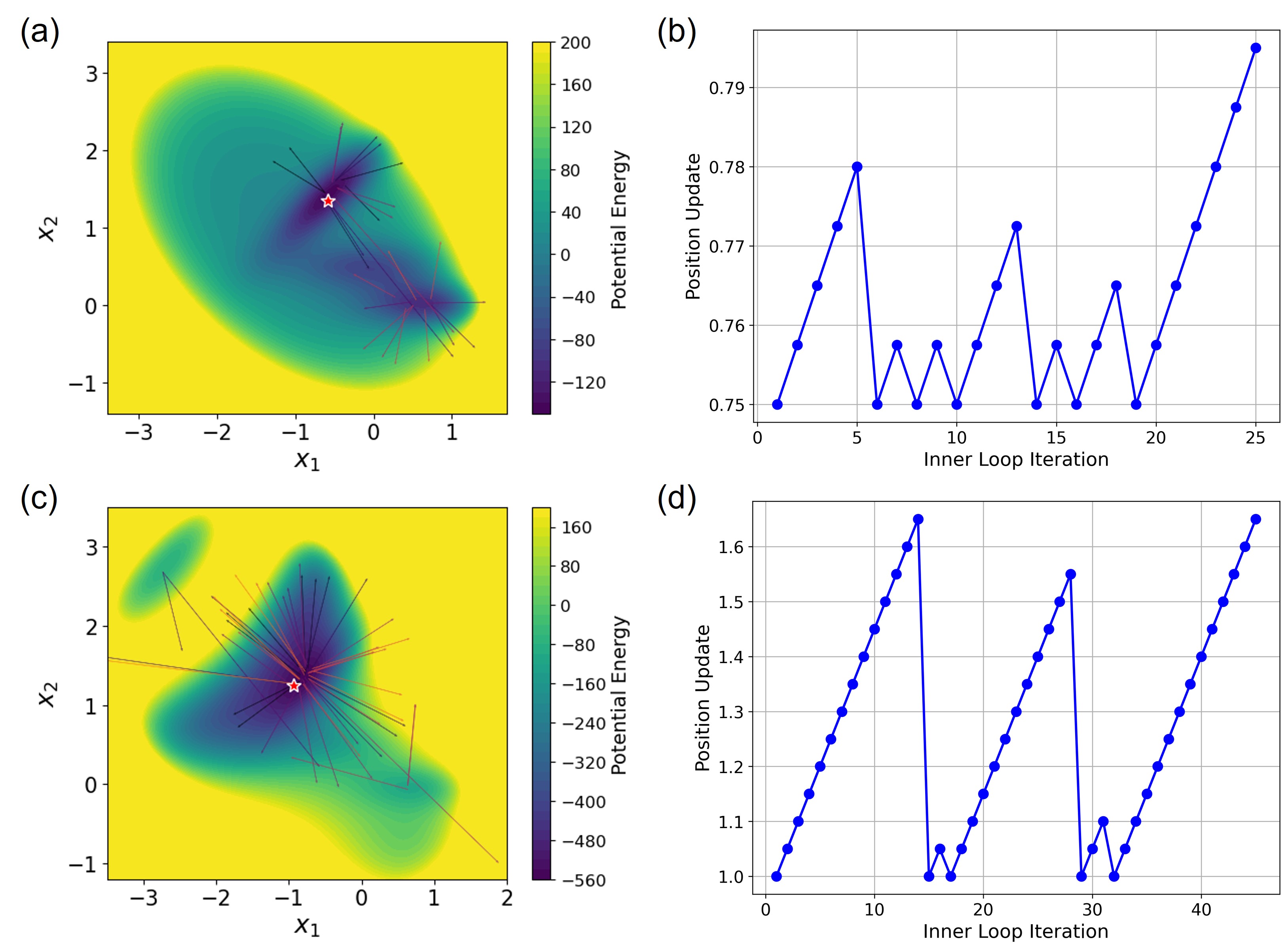} 
\caption{Simulation starting point updates. Upper panels display results from M\"uller potential, while lower panels feature results from modified M\"uller potential. Panels \textbf{a} and \textbf{c} (left) demonstrate updates from the end of one run to the start of the next, marked by arrows. Panels \textbf{b} and \textbf{d} (right) show how update rates change across inner loop iterations and reset to zero when a new well is found. For clarity, updates from the first 5,000 frames are depicted.}
\label{fig:mullers_update}
\end{figure*}

The process of escaping the deepest well involves iterative re-initiation of inner loop runs, ultimately leading to the discovery of a nearby metastable state. This method involves updating the initial positions for the simulations within the inner iteration loop. The updates are directed away from the deep potential wells, aligning with the direction of the negative observation density gradient, as depicted in Figure \ref{fig:mullers_update} (a) and (c). Moreover, the magnitude of these updates progressively increases until the inner loop runs within the inner loop successfully pinpoint a new potential well, as demonstrated in Figure \ref{fig:mullers_update} (b) and (d). Although the update function is set as a linear function in this study, exploring other types of incremental adjustments could be beneficial.

\subsection{Initialization Sensitivity}

Ordinary molecular simulations often become trapped within deep potential wells, making the simulations sensitive to initial starting points. However, the GradNav algorithm enhances the exploration of energy landscapes, thereby diminishing this sensitivity to the initial setups. Consequently, gaining holistic insights into the energy surface becomes feasible, irrespective of where the simulation begins. To evaluate this reduced sensitivity, we introduce a metric named the search success initialization ratio (SSIR). This metric is calculated as the ratio of the total number of successful identifications of potential wells across iterations \(\sum_{i} N_{\text{success}}^i\) to the product of the total number of potential wells in the energy surface \(N_{\text{wells}}\) and the number of initializations \(N_{\text{init}}\), across the energy surface:

\begin{equation}
\text{SSIR [\%]} = \frac{\sum_{i} N_{\text{success}}^i}{N_{\text{wells}} \times N_{\text{init}}}
\end{equation}

\begin{figure*}[htbp] %[h!]
\centering
\includegraphics[width=0.8\textwidth]{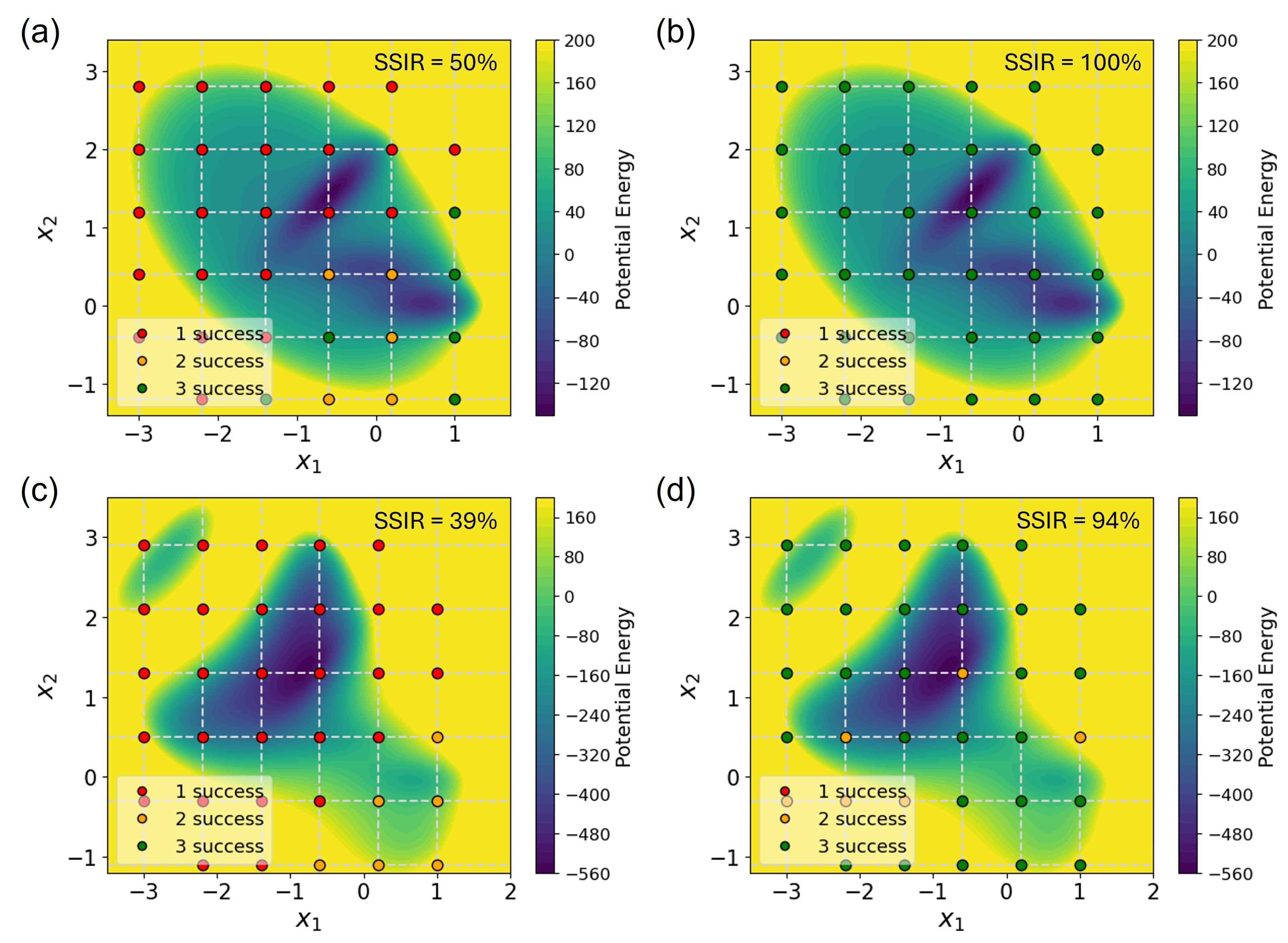} 
\caption{Count of successful potential well identifications by initial position. The left panels, \textbf{a} and \textbf{c}, depict the results obtained using Langevin dynamics (LD). In contrast, the right panels, \textbf{b} and \textbf{d}, illustrate the outcomes derived from GradNav.}
\label{fig:mullers_ssir}
\end{figure*}

The iteration of initial points is conducted using a grid spacing of 0.8, with each iteration spanning 10,000 frames. The number of successful potential well identifications for each initial point is denoted by distinct colors in Figure \ref{fig:mullers_ssir}. For Langevin dynamics simulations, the SSIR values are calculated to be 50\% for the Müller potential and 39\% for the modified Müller potential energy surfaces. The implementation of the GradNav algorithm leads to a substantial increase in SSIR values, reaching 100\% and 94\% for the Müller and modified Müller potential energy surfaces, respectively. This significant improvement underscores the GradNav algorithm's ability to lessen the dependency on initial positioning, thereby streamlining the need for comprehensive initialization procedures.

\subsection{Potential Energy Surface Reconstruction}

In specific potential energy surfaces, the distribution of trajectories adheres to the Boltzmann distribution. This distribution is a fundamental concept in statistical mechanics that describes the probability of a system being in a particular state as a function of that state's energy. The Boltzmann distribution can be expressed as follows\cite{boltzmann1, boltzmann2}:

\begin{equation}
p_i \propto \exp \left( -\frac{\varepsilon_i}{k_B T} \right) \Leftrightarrow \frac{\varepsilon_i}{k_B T} \propto -\ln p_i 
\label{eq:boltzmann}
\end{equation}

Here, \(p_i\) denotes the probability of the system being in state \(i\), \(\varepsilon_i\) represents the energy of state \(i\), \(k_B\) is the Boltzmann constant, and \(T\) symbolizes the thermodynamic temperature. In the context of Langevin dynamics simulations involving a single particle, the ``system" in question refers to this individual particle.

\begin{figure*}[htbp] %[h!]
\centering
\includegraphics[width=0.8\textwidth]{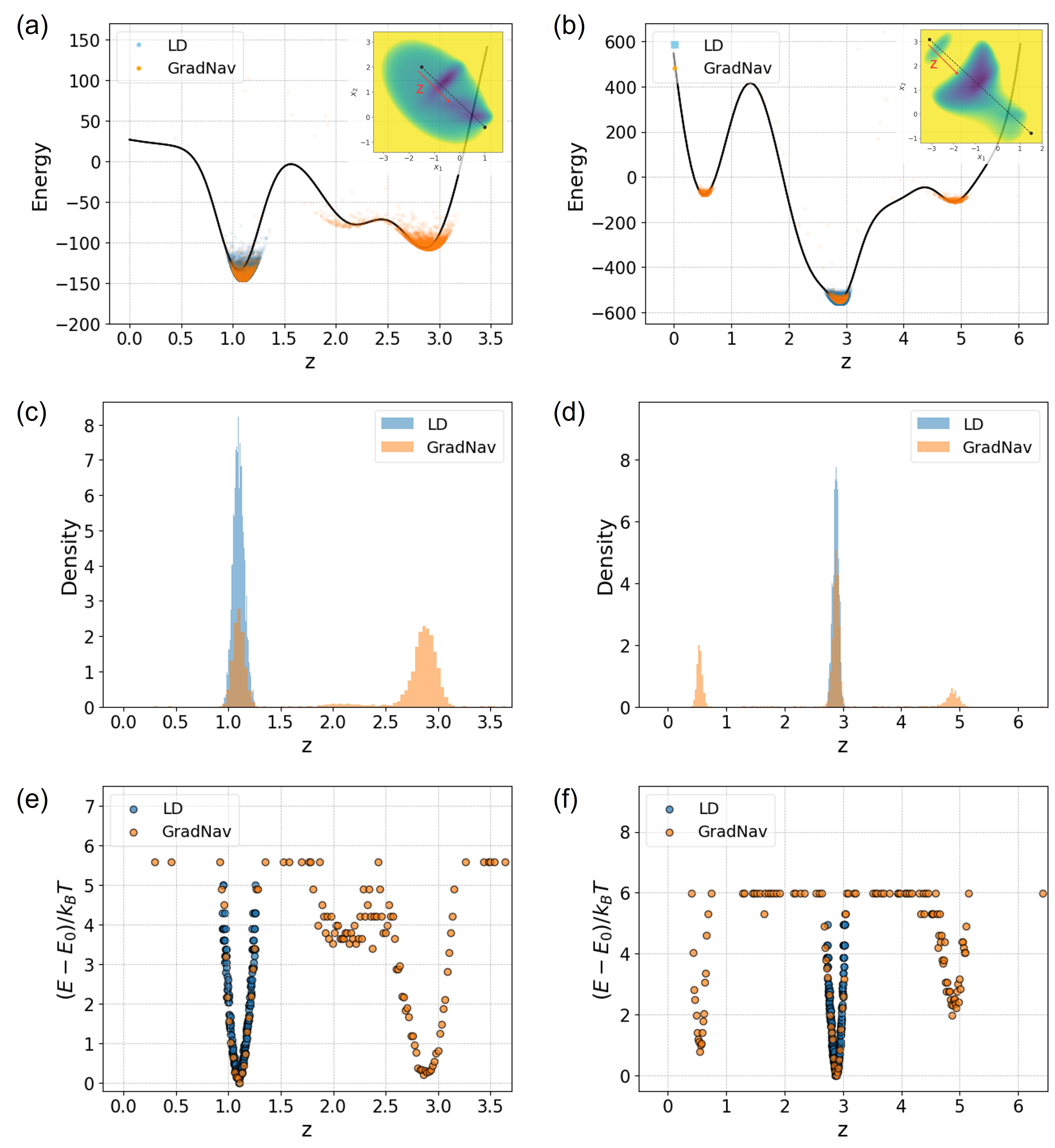} 
\caption{Boltzmann distribution of trajectories. The left panels depict results from the Müller potential and the right panels from the modified Müller potential. Panels \textbf{a} and \textbf{b} show trajectories on the energy cross-section. Panels \textbf{c} and \textbf{d} present the distribution histogram of trajectories. Panels \textbf{e} and \textbf{f} illustrate energy estimates derived using the Boltzmann distribution equation. The energy of state \(i\), denoted as \(\varepsilon_i\), is represented by \(E(\mathbf{X})\) within the energy surface.
}
\label{fig:mullers_energy}
\end{figure*}

Utilizing this relationship allows for the estimation of the energy based on the probabilities extracted from the particle's trajectory data\cite{boltzmann_reconstruct}. Figure \ref{fig:mullers_energy} displays the estimated energy surfaces for both the Müller and modified Müller potentials: Müller potential results are shown in the left panels, while the modified Müller potential results appear in the right panels. The top panels, (a) and (b), display trajectory points on the energy curve extracted from cross-sections, while the cross-section creation is detailed in the inner panels. Panels (c) and (d) feature a density histogram of these points. Probabilities from the histogram are converted into energy estimates via Equation \ref{eq:boltzmann}, with adjustments made to match the lowest energy state, \(E_0\).

Trajectories from Langevin dynamics simulations in both potential energy surfaces often find themselves localized within deep potential wells, confining the distribution of points to these regions. In contrast, GradNav simulations are capable of exploring all potential wells, offering a comprehensive view of the energy surface. As a result, energy curve estimates derived from Langevin dynamics simulation data only capture a single, deep potential well where the trajectories are trapped. On the other hand, GradNav provides a holistic reconstruction of the energy curve across the entire cross-sections for both potentials. This underscores the necessity for an expansive exploration to accurately estimate the energy surface from observed data. Additionally, capturing trajectories that span various metastable states is essential for effectively training deep learning models tasked with meticulously mapping these trajectories onto the latent space \cite{bg_noe}.

\subsection{Pseudo Molecular Dynamics}

\begin{figure*}[htbp] %[h!]
\centering
\includegraphics[width=0.8\textwidth]{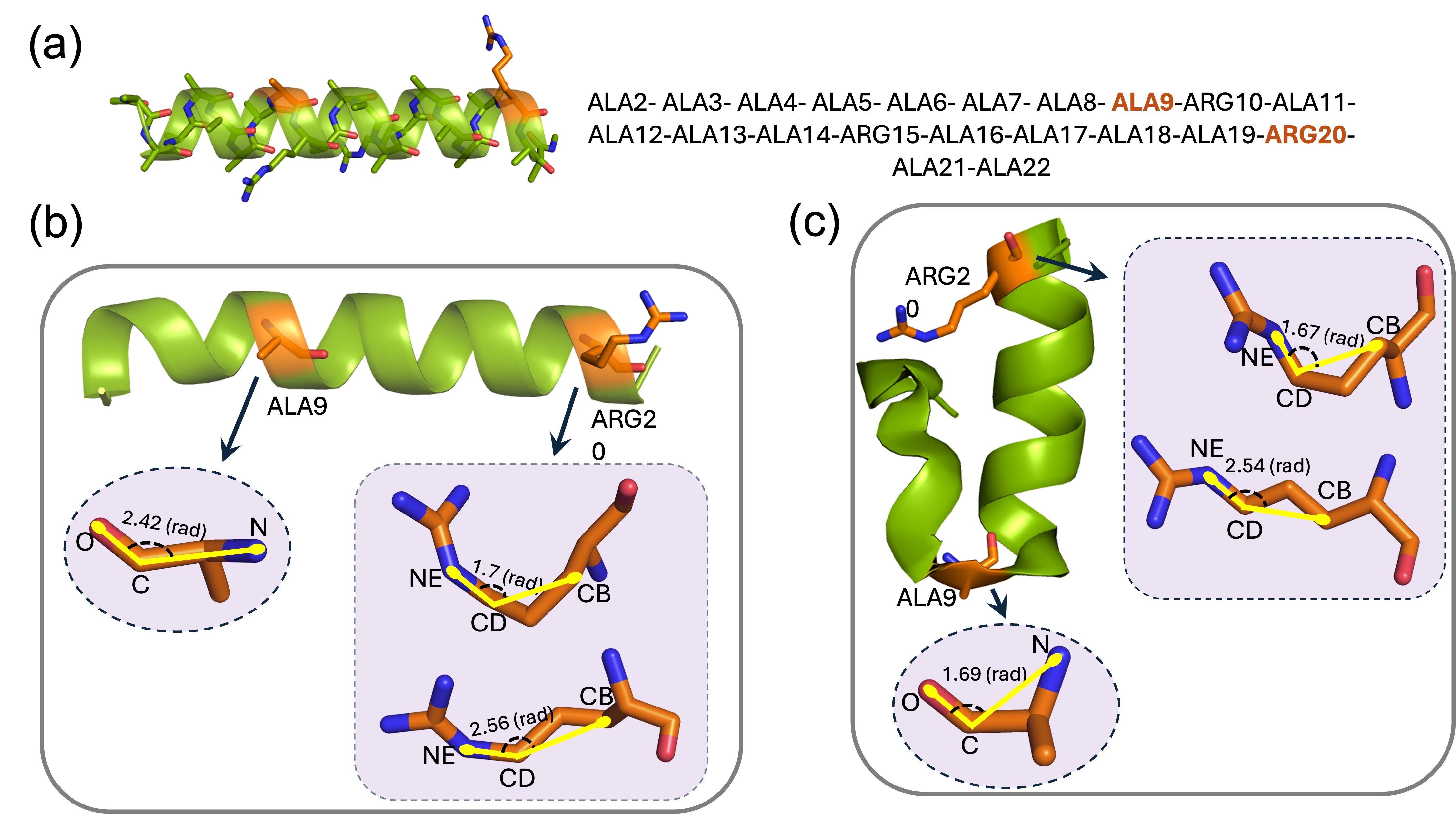} 
\caption{Topology visualization of Fs-Peptide. \textbf{a} Amino acid sequence in Fs-Peptide protein. Unfolded \textbf{b} and folded \textbf{c} states of Fs-Peptide protein with dynamics of ALA9 (stable switch) and ARG20 (unstable switch) amino acids within the protein.}
\label{fig:fs-peptide}
\end{figure*}

To evaluate the GradNav algorithm with a real-world system, we use molecular dynamics trajectories from the Fs-Peptide protein, building upon earlier demonstrations with model systems. For this validation, we employ an existing trajectory dataset in a manner akin to pseudo molecular dynamics simulations, bypassing the execution of actual molecular dynamics simulations. This dataset consists of 28 trajectories, comprising a total of 280,000 frames, illustrating a variety of behaviors and topologies\cite{fs-peptide}. The topology of the Fs-Peptide protein is visualized in Figure \ref{fig:fs-peptide}. While certain trajectories demonstrate protein folding, others lack this characteristic behavior (see Supporting Information S2). Detailed information on the simulation settings and dataset specifics is provided in the Methods section.

The pseudo molecular dynamics process initiates with the selection of a
single trajectory, starting from its initial frame. The simulation's progression is emulated by collecting subsequent frames as per the designated number in either an outer or inner loop run. A new potential starting point on the energy surface is then proposed based on the last frame and the GradNav algorithm's update equation. Points within a cutoff radius of 0.02 from the last frame are identified, from which one is randomly chosen as the new starting point. If no points are located within this radius, the closest point is selected instead. Figure S2 in Supporting Information illustrates both the initial points tentatively proposed by the update rule and the actual initial points selected from the dataset. This step is followed by the continuation of the process for a predetermined number of frames, defined as the frame count for each loop, from this new starting point. 

This methodology mirrors the initialization of the starting point based solely on identified collective variables, which serve as axes in the surface. Given that only collective variables are known for the new point, it may not be feasible to specify a unique molecular system solely on these collective variables. Our approach, therefore, involves random initialization based on the updated collective variables, notably the \ce{O}-\ce{C}-\ce{N} angle in the alanine 9 (ALA9) residue and the \ce{CB}-\ce{CD}-\ce{NE} angle in the arginine 20 (ARG20) residue. The random selection from candidate points adheres to the generation of a new initial structure with constraints on these two angles, indicating that the simulation may re-initiate from any structure that aligns with the newly updated angles for ALA9 and ARG20. With the dataset encompassing 280,000 frames across 28 trajectories, it ensures structural diversity.

\subsection{Validation with Fs-Peptide}

\begin{figure*}[htbp] %[h!]
\centering
\includegraphics[width=0.8\textwidth]{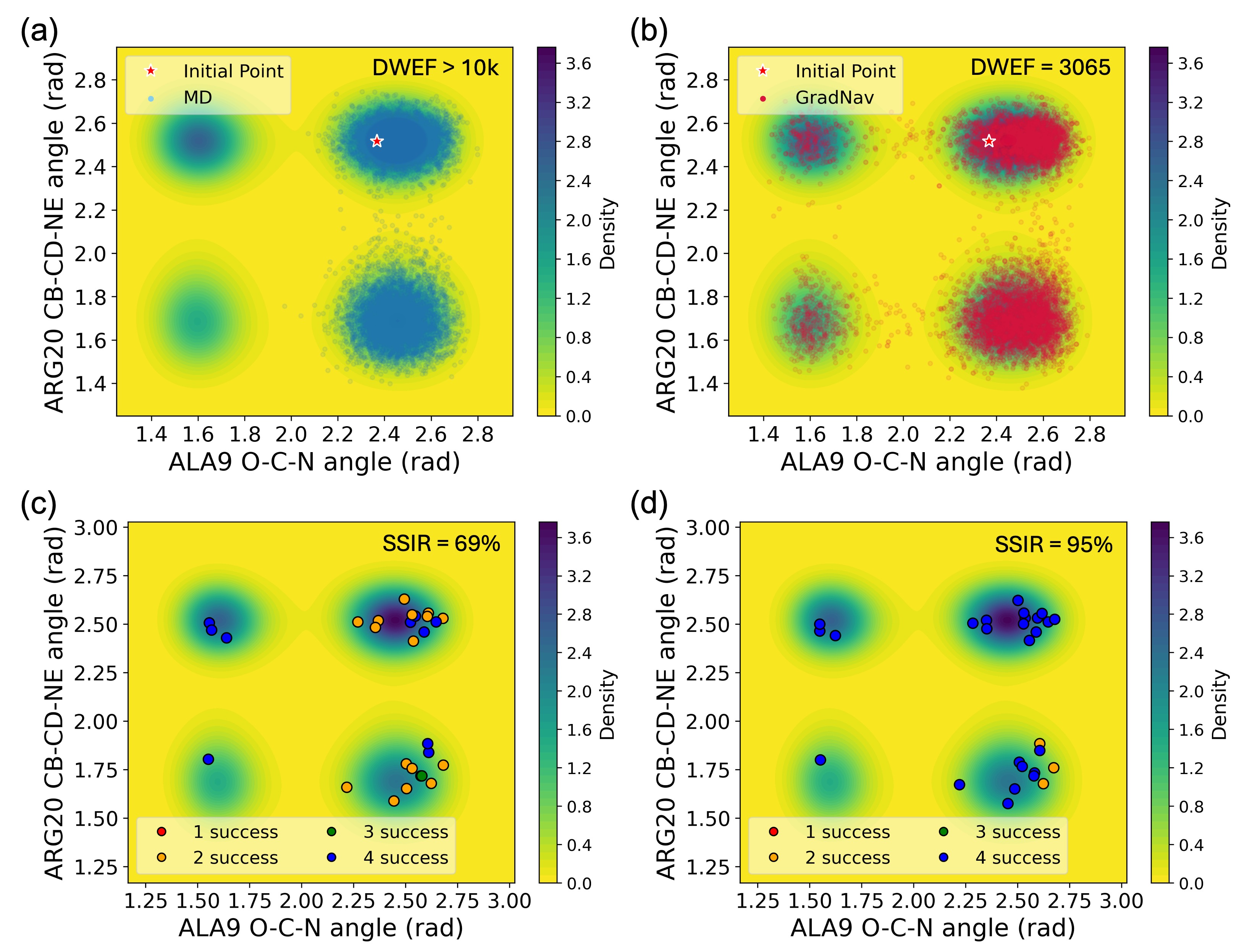} 
\caption{Trajectories across the energy surface with initial points and the count of successfully identified metastable states. Panels \textbf{a} and \textbf{c} show the molecular dynamics (MD) trajectory, labeled as number 4 in the dataset, while panels \textbf{b} and \textbf{d} display outcomes from GradNav, starting from the same points as MD.}
\label{fig:peptide_traj_ssir}
\end{figure*}

We demonstrate the ability to escape deep potential wells and the reduced sensitivity to the initial point using Fs-Peptide trajectories, as shown in Figure \ref{fig:peptide_traj_ssir}. A density map in the background, derived from a single trajectory (specifically, number 15 in the dataset), showcases stable protein folding behavior. While this map, originating from one trajectory, may not fully capture the comprehensive energy surface, it offers a preliminary estimation. Likewise, in real-world scenarios, acquiring a complete understanding of the energy surface is often infeasible. The trajectories are analyzed using two collective variables: the angle formed within \ce{O}-\ce{C}-\ce{N} atoms in the ALA9 residue and the \ce{CB}-\ce{CD}-\ce{NE} angle in the ARG20 residue, where the former being correlated to the Fs-Peptide's folding behavior\cite{mollaei2023unveiling} (see Figure \ref{fig:fs-peptide}). The escape from a deep potential well is marked by a notable shift in the ALA9 angle, transitioning from approximately 2.5 rad to about 1.6 rad. On the other hand, the identification of all metastable states in SSIR remains the same as in previous cases, involving probing all density wells within the energy surface. 
%The topology of the Fs-Peptide protein is visualized in Figure S2 of the Supporting Information.

We assess the performance of GradNav against a specific molecular dynamics trajectory that is unable to escape a region around an ALA9 angle ranging from 2.2 rad to 2.8 rad. To ensure a fair comparison, GradNav initiates from the same starting point and follows the trajectory of the molecular dynamics simulation. Figures \ref{fig:peptide_traj_ssir} (a) and (b) display the trajectory points within the energy surface for molecular dynamics and GradNav, respectively. While the molecular dynamics simulation remains trapped within the vicinity of an ALA9 angle of 2.5 rad throughout 10,000 frames, GradNav facilitates escape from this region in just 3,065 frames.

This ability to navigate the molecular space extensively with GradNav is further evidenced by its reduced sensitivity to the initial setup. Figures \ref{fig:peptide_traj_ssir} (c) and (d) illustrate the number of successful identifications of metastable states from each initial point, represented by different colors. In this analysis, GradNav operates in a pseudo molecular dynamics fashion, applied to all initial points across the 28 trajectories in the dataset. It preserves a total frame count of 10,000, which mirrors the frame count per molecular dynamics trajectory. When relying solely on molecular dynamics simulations, the SSIR is 69\%. This rate increases to 95\% with the application of GradNav, highlighting its diminished dependence on the initial setup.

\begin{figure*}[htbp] %[h!]
\centering
\includegraphics[width=0.9\textwidth]{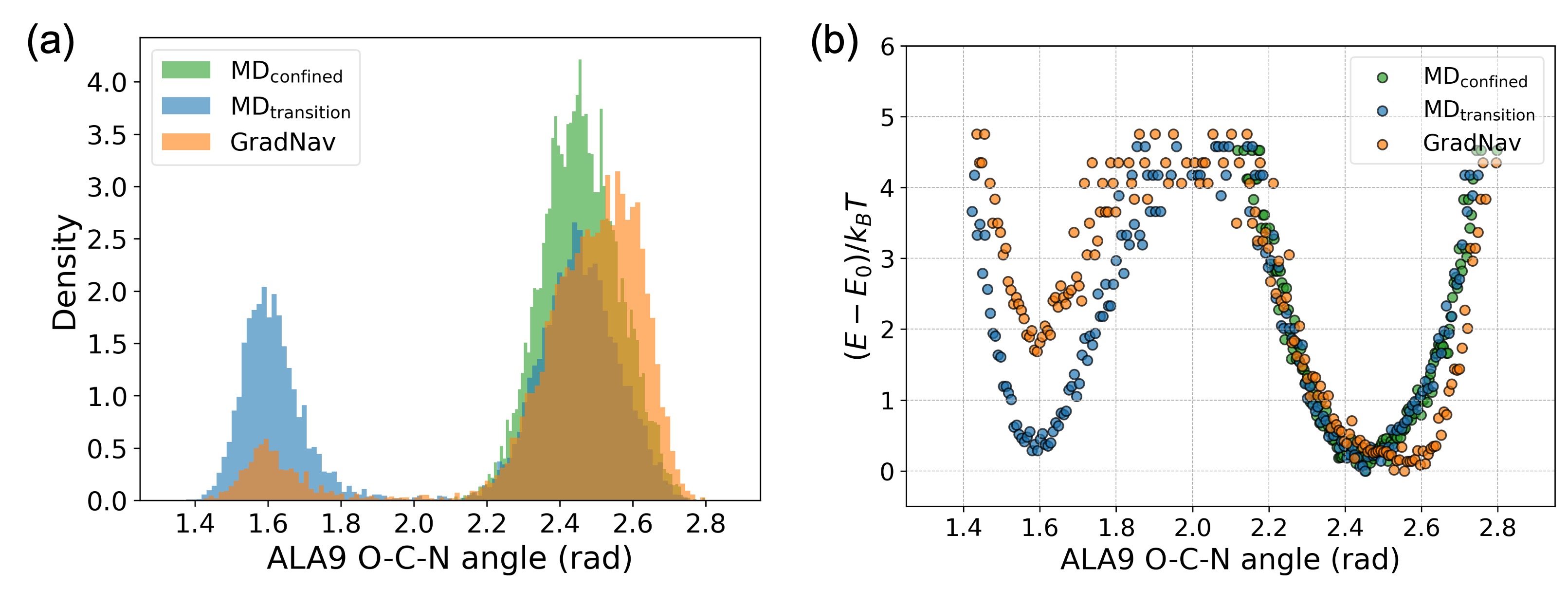} 
\caption{Estimations of energy derived from the distribution of trajectories. Panel \textbf{a} presents a histogram of trajectory distribution. Panel \textbf{b} illustrates the reconstructed energy estimations based on the trajectory histogram. MD$_\text{confined}$ corresponds to trajectory number 4, and MD$_\text{transition}$ to trajectory number 15.}
\label{fig:peptide_energy}
\end{figure*}

% The more metastable identified by the trajectory, the reconstructed energy estimates become more comprehensive. 

After obtaining the trajectory across the comprehensive energy surface, the energy curve can be successfully reconstructed. The molecular dynamics trajectory, denoted as MD$_\text{confined}$ in Figure \ref{fig:peptide_energy}, remains confined within the vicinity of an ALA9 angle of 2.5 rad. In contrast, the molecular dynamics simulation labeled as MD$_\text{transition}$ demonstrates a transition between two regions of ALA9 angle, exhibiting two peaks in the distribution. Figure \ref{fig:peptide_energy}(b) shows that while the trajectory trapped in a potential well fails to effectively reconstruct the energy estimate curve, the comprehensive trajectory that explores both potential wells enables successful reconstruction. The trajectory derived from GradNav accurately captures both potential wells. However, there is a discrepancy in the potential well depths between the molecular dynamics and GradNav trajectories. Upon identifying all potential wells, their depths can be accurately determined through targeted molecular simulations in those specific regions.

\begin{figure*}[htbp] %[h!]
\centering
\includegraphics[width=0.75\textwidth]{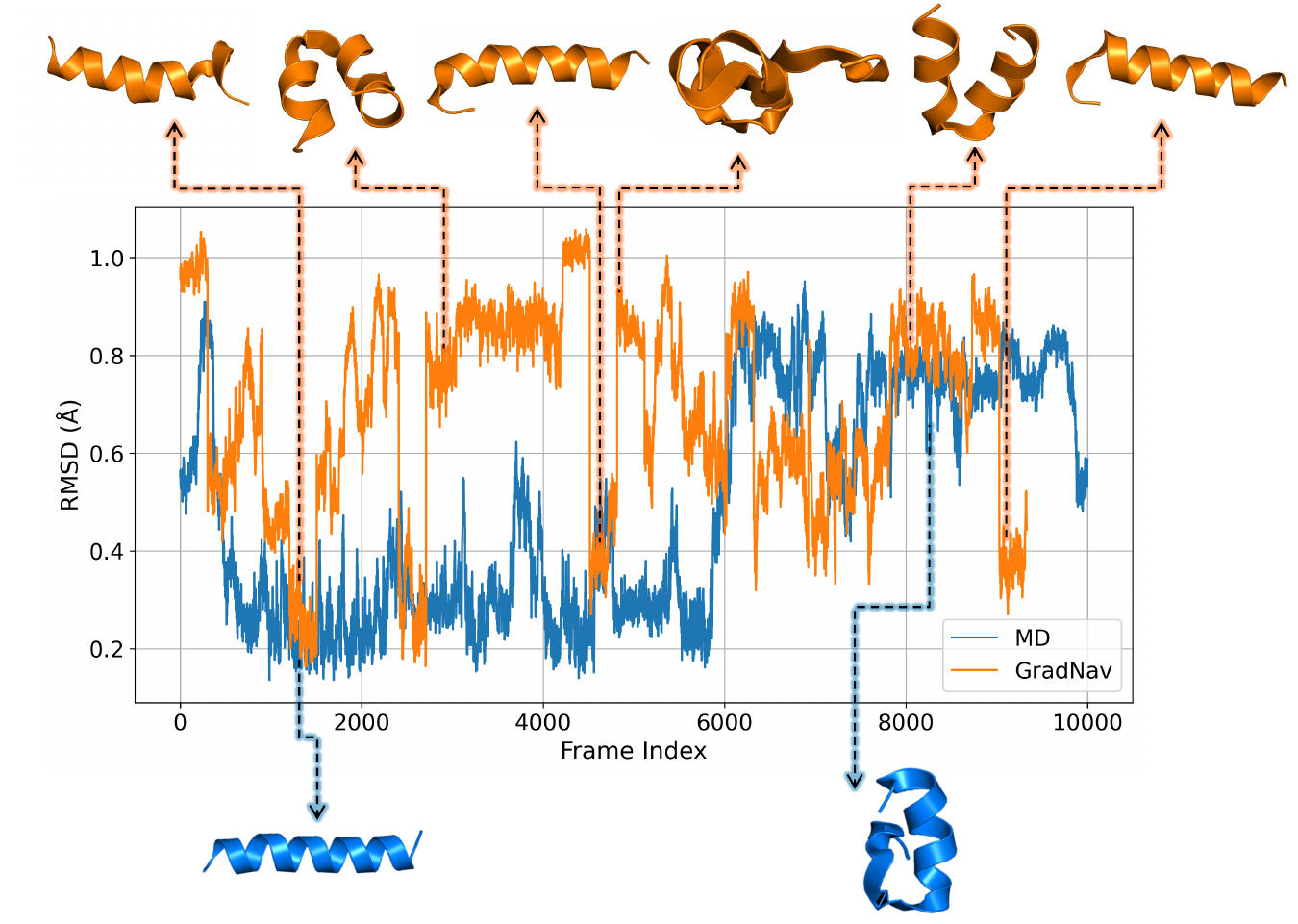} 
\caption{Comparison of root mean square deviation between MD simulation and GradNav algorithm trajectories, showcasing structural variation over time. MD trajectory here is the same as MD$_\text{transition}$.}
\label{fig:peptide_rmsd}
\end{figure*}

A key focus of molecular dynamics simulations of proteins is understanding their folding behavior, which significantly influences the protein's functional characteristics. Figure \ref{fig:peptide_rmsd} illustrates the root mean square deviation (RMSD) of each frame relative to a reference frame in an unfolded state, providing insight into the folding process. For example, a specific molecular dynamics trajectory demonstrates a stable transition towards a folded state, as evidenced by a jump in the RMSD plot. The trajectory produced by the GradNav algorithm exhibits frequent alternations between unfolded and folded states. This behavior is expected, given that GradNav generates its trajectory through multiple simulation restarts, offering a perspective not of a continuous path but rather an exploration across a broad region of the energy surface. This exploration reveals multiple potential folding states for further examination. GradNav enables the identification of various folded and unfolded states of the Fs-Peptide and provides initial conditions for each re-initiation of the simulations. This underscores its utility in probing the protein's conformational landscape, making GradNav an effective tool for gathering candidate conformations for a detailed analysis of folding dynamics.

\section{Conclusion}

We have introduced the GradNav algorithm, designed to traverse the energy surface by iteratively running short segments of molecular simulation and updating the initial (seeding) points towards the gradient of observation density. This approach systematically guides the simulation into previously unexplored regions. Furthermore, we proposed two evaluative metrics: DWEF and SSIR. DWEF measures the trajectory's efficiency in escaping deep potential wells, with lower DWEF values indicating superior escape capabilities. Meanwhile, SSIR assesses the trajectory's dependency on the very first starting points, with higher SSIR values signifying a reduced sensitivity to these starting locations.

We have implemented the GradNav algorithm in Langevin dynamics simulations of a single particle within both the Müller potential and modified Müller potential energy surfaces, as well as in molecular dynamics simulations of the Fs-Peptide protein. Our analysis demonstrates that GradNav significantly outperforms unenhanced molecular simulation approaches by efficiently navigating out of deep energy wells, as evidenced by lower DWEF values. This capability to explore more effectively reduces the dependency on the initial starting point of the simulation, confirmed by GradNav's higher SSIR values. GradNav's systematic exploration strategy thus leads to a more precise assessment of the energy surface through the distribution of trajectories it generates, improving the accuracy of energy estimates.

% Given that GradNav conducts multiple simulation runs to probe adjacent metastable states, the resulting trajectory collection should be interpreted as a series of discrete trajectories initiated from various points, rather than a single continuous trajectory. When applied to the Fs-Peptide molecular dynamics simulation, GradNav identified multiple transitions between the folded and unfolded states. These transitions should be considered as potential topologies and starting points for further detailed investigations.

% The potential of GradNav extends to its integration with machine learning techniques. Currently, machine learning models are adept at modeling atomic systems into latent spaces. For instance, autoencoders have been successful in mapping the Brownian dynamics trajectories of a 2D energy landscape into latent space\cite{Ramil2022}. Similarly, flow-based Boltzmann generators have effectively mapped the structure of the bovine pancreatic trypsin inhibitor (BPTI) protein into latent space while preserving distributional information from the real space\cite{bg_noe}. This preservation suggests that areas densely populated in real space should correspond to dense regions in latent space. 

Building on the application of GradNav to known collective variables, we anticipate that the algorithm will demonstrate similar proficiency in moving from densely populated to unexplored areas within the latent space of machine learning models. Successful machine learning modeling is expected to maintain the correspondence between distributions in actual physical space and their representations in the model's latent space. Such a strategy promises to improve the exploration of complex molecular systems by leveraging the advantages of both simulation algorithms and machine learning models.

\section{Methods}

\subsection{Langevin Dynamics}

Langevin dynamics is a stochastic simulation technique used to model the behavior of particles in a thermodynamic system. It extends classical molecular dynamics by including random collisions with an implicit solvent, effectively capturing the effects of thermal fluctuations. The Langevin equation integrates Newton's second law of motion with stochastic terms to describe the dynamics of particles. The motion of a particle with mass \(m\) is captured by the following equation\cite{LD1, LD2, LD3}:
\begin{equation}
m\frac{d^2\mathbf{X}}{dt^2} = -\nabla V(\mathbf{X}) - \gamma\frac{d\mathbf{X}}{dt} + \sqrt{2\gamma k_B T}\mathbf{R}(t)
\end{equation}

where \(\mathbf{X}\) denotes the position vector of the particle, \(\nabla V(\mathbf{X})\) represents the force derived from the potential energy \(V\) acting on the particle, \(\gamma\) is the friction coefficient signifying the resistance encountered by the particle due to its interaction with the solvent, \(k_B\) is the Boltzmann constant, \(T\) denotes the temperature of the system, and \(\mathbf{R}(t)\) represents the stochastic term. This stochastic term models the random forces as Gaussian white noise, ensuring that the system adheres to the fluctuation-dissipation theorem. This theorem is crucial as it guarantees the system's return to equilibrium following perturbation, thus accurately reflecting the natural behavior of particles in thermal environments.

In this study, Langevin dynamics simulations were performed using the OpenMM Python package\cite{openmm}, with the time step set to 100 fs and the total number of frames at 10,000. Simulation parameters included the potential energy \(V\), represented by two potential energy surfaces: the Müller and a modified Müller potential. The friction coefficient \(\gamma\) was set at 100 ps\(^{-1}\), the mass \(m\) was set as 1 Da, and the temperature \(T\) at 750 K.

\subsection{Model Potential Energy Surface}

In this study, we utilize the M\"uller potential along with its modified versions to define the potential energy \(V\) acting on a single particle in the Langevin dynamics simulations. The first model energy surface explored in this study is the widely studied M\"uller potential, characterized as the sum of four Gaussian functions, defined as follows:

\begin{equation}
V_{\text{MP}}(x_1, x_2) = \sum_{i=1}^{4} A_i \exp \left[ a_i(x_1 - \beta_i)^2 + b_i(x_1 - \beta_i)(x_2 - \gamma_i) + c_i(x_2 - \gamma_i)^2 \right]
\end{equation}

Here, \(A\), \(a\), \(b\), \(c\), \(\beta\), and \(\gamma\) represent parameter vectors, specifically set as \(A = [-200, -100, -170, 15]\), \(a = [-1, -1, -6.5, 0.7]\), \(b = [0, 0, 11, 0.5]\), \(c = [-10, -10, -6.5, 0.7]\), \(\beta = [1, 0, -0.5, -1]\), and \(\gamma = [0, 0.5, 1.5, 1]\), respectively. A contour map of this potential, depicted in Figure \ref{fig:mullers_traj} (a), reveals its composition: two primary wells and an intermediate, shallower well, with their center coordinates estimated to be (-0.55, 1.45), (0.65, 0.02), and (-0.1, 0.45), respectively.

To enhance the evaluation of our algorithm across a broader range of scenarios, we introduce a variation to the M\"uller potential by incorporating an additional term, \(V_{\text{add}}\). This alteration results in the modified M\"uller potential, which is articulated as follows\cite{silvia2017}:

\begin{subequations}
\begin{align}
V_{\text{MMP}}(x_1, x_2) &= V_{\text{MP}}(x_1, x_2) + V_{\text{add}}(x_1, x_2), \\
V_{\text{add}}(x_1, x_2) &= A_5 \sin(x_1 x_2) \exp \left[ a_5 (x_1 - \beta_5)^2 + c_5 (x_2 - \gamma_5)^2 \right]
\end{align}
\end{subequations}

where the parameters \(A_5\), \(a_5\), \(c_5\), \(\beta_5\), and \(\gamma_5\) are set to 500, -0.1, -0.1, -0.56, and 1.44, respectively.

This modification introduces a more complex challenge for the simulation in exploring all potential wells, as the additional term creates a deeper valley, further distancing two shallow metastable states. Consequently, for a comprehensive exploration of all minima, the simulation must successfully navigate out of the deep valley in both directions: towards the upper right and the lower left.

\subsection{Fs-Peptide Molecular Dynamics}

In this study, we investigated the dynamics of Fs-peptide protein (ACE-A\_5(AAARA)\_3A-NME), which comprises 28 molecular dynamics trajectories, each 500 ns in length.\cite{fs-peptide}. These trajectories are saved at a time interval of 50 ps, resulting in an aggregate sampling of 14 $\mu$s. These molecular dynamics simulations provide insights into the dynamic behavior of the protein, allowing us to observe how the peptide folds and unfolds over time. The simulations were performed using the AMBER99SB-ILDN force field, a commonly used force field for simulating biomolecular systems\cite{lindorff2010improved, weiner1986all}. Additionally, the simulations utilized the Generalized Born Surface Area (GBSA) implicit solvent model with Onufriev-Bashford-Case (OBC) parameters\cite{onufriev2004exploring}. Implicit solvent models approximate the effects of solvent without explicitly representing water molecules, making simulations computationally more efficient. The simulations were conducted at a temperature of 300 K, which is typical for simulating biological systems under physiological conditions. The simulations started from randomly sampled conformations obtained from an initial 400 K unfolded simulation, likely providing a diverse set of starting conformations to explore the folding behavior of the peptide. The simulations were performed using OpenMM 6.0.1, a versatile molecular dynamics simulation toolkit that offers high performance and flexibility for simulating biomolecular systems.

% \begin{acknowledgement}

% Please use ``The authors thank \ldots'' rather than ``The
% authors would like to thank \ldots''.

% The author thanks Mats Dahlgren for version one of \textsf{achemso},
% and Donald Arseneau for the code taken from \textsf{cite} to move
% citations after punctuation. Many users have provided feedback on the
% class, which is reflected in all of the different demonstrations
% shown in this document.\cite{ramil2022}

% \end{acknowledgement}
% \section*{Technology Use Disclosure}
% ChatGPT was used to help prepare the preprint version of this manuscript, specifically for grammar and typo corrections. All information in this manuscript has been read, corrected, and verified by all authors.

\begin{suppinfo}

The supplementary materials encompass the following elements: outer loop trajectory, Fs-Peptide protein, and update of initial points in Fs-Peptide trajectories.

\end{suppinfo}

\section*{Data Availability Statement}
The molecular dynamics trajectories of Fs-Peptide are available at: \url{https://figshare.com/articles/dataset/Fs_MD_Trajectories/1030363?file=1502287}. 

\section*{Code Availability Statement}
The Python code employed in this study is available on GitHub at the following link: \url{https://github.com/hoon-ock/landscape-search}.

\bibliography{reference}

\setcounter{table}{0}
\renewcommand{\thetable}{S\arabic{table}}
\setcounter{figure}{0}
\renewcommand{\thefigure}{S\arabic{figure}}
\setcounter{equation}{0}
\renewcommand{\theequation}{S\arabic{equation}}%%%%%%%%%%%%%%%%%%%%%%%%%%%%%%%%%%%%%%%%%%%%%%%%%%%%%%%%%%%%%%%%%%%%%
\newpage
\maketitle
\section{Supplementary Information}
% \SectionNumbersOn
% \begin{document}

% \tableofcontents

\section{Outer Loop Trajectory}

\begin{figure*}[htbp] %[h!]
\centering
\includegraphics[width=0.8\textwidth]{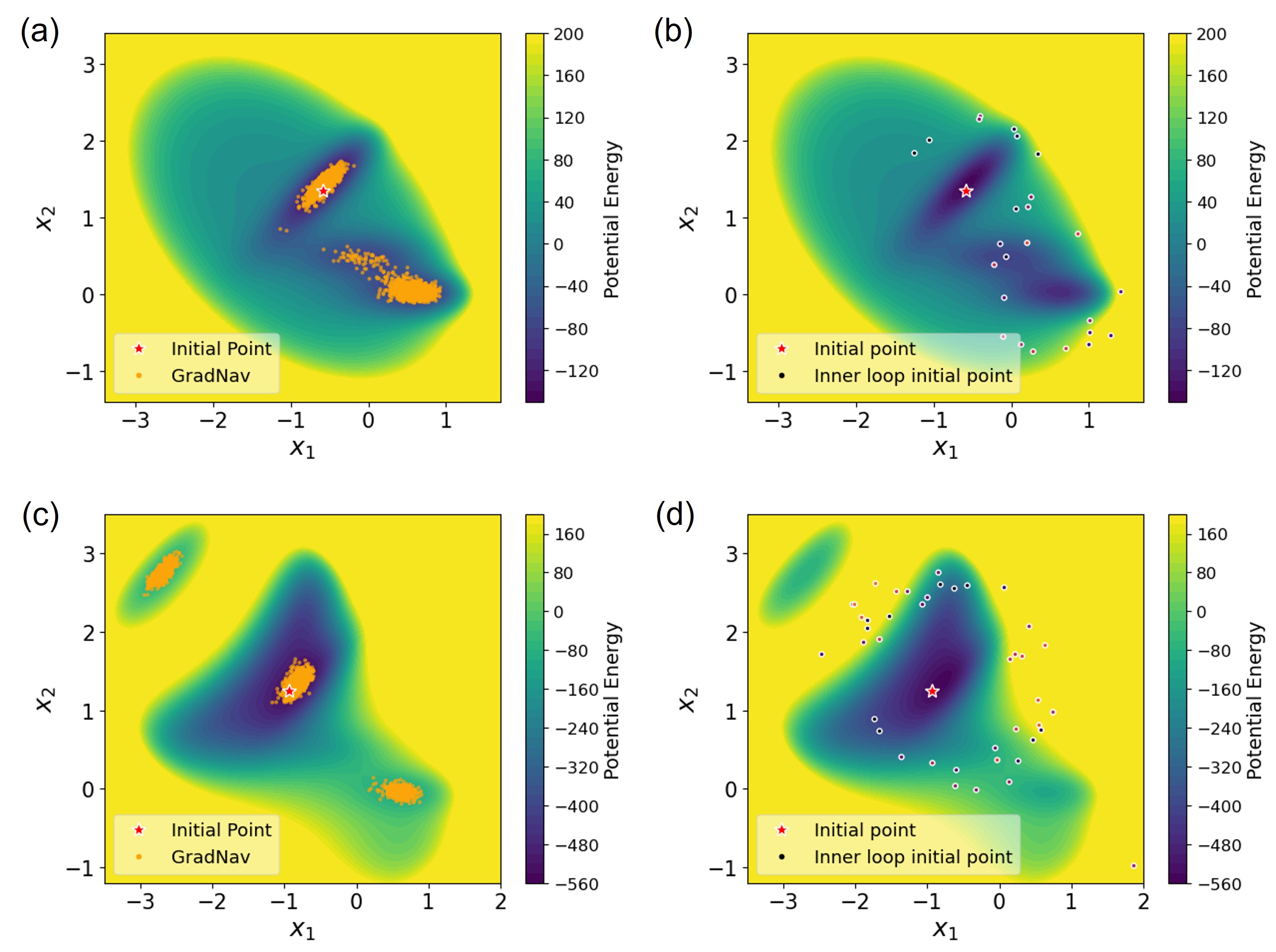} 
\caption{Trajectories from outer loop Simulations. The upper panels \textbf{a} and \textbf{b} showcase results from simulations using the Müller potential, while the lower panels \textbf{c} and \textbf{d} depict outcomes from the modified Müller potential.}
\label{fig:si_outer}
\end{figure*}

Each trajectory generated during an iteration of the outer loop simulation can be considered a reliable segment of simulation, assuming the frame duration is appropriately set. The left panels of Figure \ref{fig:si_outer} display trajectories exclusively obtained from outer loop runs. Upon examining the right panels in Figure 2, one observes trajectories more densely localized within potential wells. The majority of the dispersed points located in the high energy area of the right panels in Figure 2 correspond to the initial positions for each run of the inner loop, as depicted in the right panels of Figure \ref{fig:si_outer}.

\newpage
\section{Fs-Peptide Protein}

In one of our studies\cite{mollaei2023unveiling}, we investigated the dynamics and conformational states of individual amino acids within proteins using a combination of molecular dynamics (MD) simulations and machine learning (ML) techniques. Specifically, we focused on identifying residues that switch between two distinct angular states, classifying them as either stable switch (ALA9 in Figure 6) or unstable switch (ARG20 in Figure 6) residues, and evaluating their contribution to the overall properties of proteins. For instance, in the Fs peptide protein, we found that the Root Mean Square Deviation (RMSD) feature, indicative of protein folding, strongly correlates with the dynamics of ALA9 amino acid. It suggests that the ALA9's dynamics may contribute to the folding process of the protein. However, the ARG20 jumps between two distinct angular states regardless of the conformational states of the protein. As Figure 6 shows, the ALA9 residue is stable in a single angular state when the protein is unfolded (Figure 6b) and suddenly switches to another state once the RMSD increases (Figure 6c). While ARG20 transitions between the two angular states irrespective of the protein's conformational states.

% \begin{figure*}[htbp] %[h!]
% \centering
% \includegraphics[width=0.8\textwidth]{figures/FS.pdf} 
% \caption{(a) Amino acid sequence in Fs peptide protein. Unfolded (b) and folded (c) states of Fs peptide protein with dynamics of ALA9 (stable switch) and ARG20 (unstable switch) amino acids within the protein.}
% \label{fig:fs}
% \end{figure*}

\newpage
\section{Update of Initial Points in Fs-Peptide Trajectories}

\begin{figure*}[htbp] %[h!]
\centering
\includegraphics[width=0.6\textwidth]{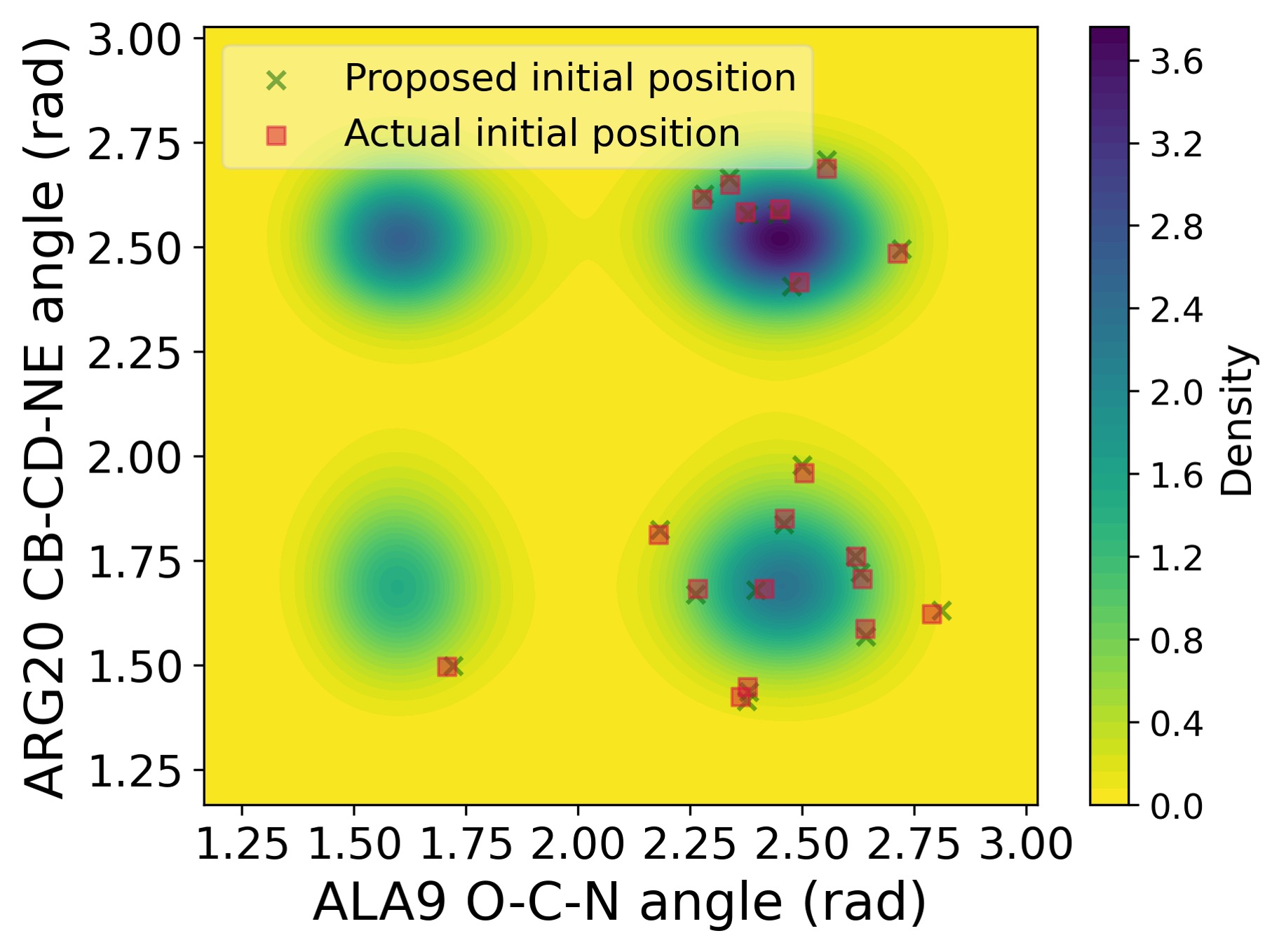} 
\caption{Proposed vs. actual initial points. The proposed initial points are updated using the GradNav algorithm's update rule, while the actual initial points are selected from the trajectory dataset based on a cutoff radius of 0.02.}
\label{fig:si_init}
\end{figure*}

We assess the implementation of GradNav in Fs-Peptide molecular dynamics simulations through a pseudo molecular dynamics approach. In this method, a new starting point is proposed, followed by the selection of an actual starting point located within a cutoff radius from the newly proposed point. Given that the dataset includes 280,000 trajectory frames, thoroughly covering the molecular space with diverse topologies, we consistently identified points within the cutoff radius, as depicted in Figure \ref{fig:si_init}.

\newpage
% \end{document}
\end{document}